\documentclass[conference]{IEEEtran}
\usepackage{cite}
\usepackage{amsmath,amssymb,amsfonts}
\usepackage{algorithmic}
\usepackage{graphicx}
\usepackage{textcomp}
\usepackage{xcolor}
\usepackage{booktabs}

\begin{document}

\title{Image-Hashing-Based Anomaly Detection for Privacy-Preserving Online Proctoring\\
\thanks{Identify applicable funding agency here. If none, delete this.}
}

\author{\IEEEauthorblockN{Waheeb Yaqub}
\IEEEauthorblockA{\textit{CHAI Lab - School of Computer Science} \\
\textit{The University of Sydney}\\
Sydney, Australia \\
waheeb.faizmohammad@sydney.edu.au}
\and
\IEEEauthorblockN{Manoranjan Mohanty}
\IEEEauthorblockA{\textit{Center for Forensic Science} \\
\textit{University of Technology Sydney}\\
Sydney, Australia \\
manoranjan.mohanty@uts.edu.au}
\and
\IEEEauthorblockN{Basem Suleiman}
\IEEEauthorblockA{\textit{School of Computer Science} \\
\textit{The University of Sydney}\\
Sydney, Australia \\
basem.suleiman@sydney.edu.au}
}

\maketitle


\begin{abstract}
Online proctoring has become a necessity in online teaching. Video-based crowd-sourced online proctoring solutions are being used, where an exam-taking student's video is monitored by third-parties, leading to privacy concerns. In this paper, we propose a privacy-preserving online proctoring system. The proposed image-hashing-based system can detect the student's excessive face and body movement (i.e., anomalies) that is resulted when the student tries to cheat in the exam. The detection can be done  even if the student's face is blurred or masked in video frames. Experiment with an in-house dataset shows the usability of the proposed system. 
\end{abstract} 

\begin{IEEEkeywords}
online proctoring, COVID, online exams, privacy preserving, anonymisation
\end{IEEEkeywords}

\section{Introduction}
Online teaching has increased during this COVID time. Many universities (and other educational organizations) are planning to stick to the online or blended mode teaching even after the COVID. Online teaching can involve online exams, which can require live online proctoring. In online proctoring, a student can appear for the exam by switching on the front camera (e.g., web camera) of her device (which will capture the live video), and a proctor can identify any cheating attempt by monitoring the student's live exam-taking video. Proctoring a class of hundreds of students on a computer screen by a few university staffs, however, is both tedious and error-prone (unlike the in-class proctoring). Involving a large number of university staffs also seems to be infeasible as there can be multiple parallel exams. Recruiting a large number of permanent staffs is expensive for the university, especially when an exam occurs once or twice in a semester.

   \begin{figure}[ht]
    \centering
    \includegraphics[width=0.48\textwidth]{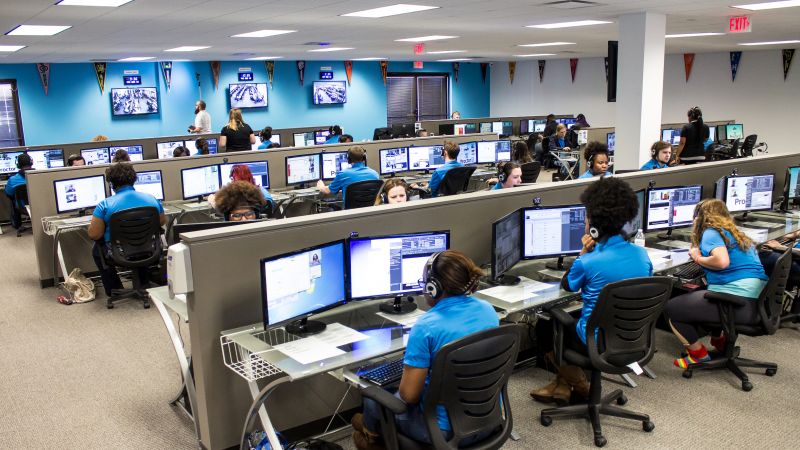}
    \caption{One of the widely used proctoring system ProctorU, where Proctors are monitoring online exam takers. The image has been taken from \cite{Onlineex65:online}.}
    \label{fig:Example}
    \end{figure}

A possible solution is to outsource the proctoring task to a third-party company. The company (which can be in a different country) can employ an adequate number of people (e.g., by crowd sourcing the proctoring task) for monitoring the students (as shown in Figure~\ref{fig:Example}). 
Although such an arrangement can address above issues to some extent, privacy is a major concern \cite{nigam2021systematic}\cite{furby2020you}. 
The availability of the student's videos (which contain faces and other personal information) to a number of casual individuals can lead to serious privacy consequences (e.g., leaking a video on a public social media platform, impersonating young students, etc.). Note that this privacy issue does not go away even if the university hires a number of  casual employees or crowd-source the proctoring task instead of outsourcing to the company \cite{Onlineex65:online}\cite{furby2020you}. 

In this paper, we propose a scheme that can help in addressing the privacy issue in online proctoring. In this scheme, a student's face is either blurred or masked in her video for protecting privacy. The blurring or masking is done in such a way that the student's excessive face and body movements can be detected. We believe that such excessive face and body movement results when the student is looking away from the computer screen for potentially taking help in the exam (i.e., trying to cheat). We call such excessive face and body movements as anomalies (with respect to normal exam taking behaviour). For detecting these anomalies in a video containing blurred/masked faces, image hashing has been used. Frames of the exam-taking video are hashed and compared to each other. When the difference in two hash values is above a threshold, it is assumed that the student has deviated from her normal exam-taking pose for attempting cheating. Such detected anomalies are then reported to humans who reconfirm any suspicious activities by looking at the blurred or masked videos. Both the tool-based anomaly detection and human reconfirmation happen at the third-party company-end. The company will then report such suspicious activities to the university, who can take further action. The university can have access to the plaintext exam-taking suspicious video.


One of the major challenge of the proposed idea is the lack of a public exam-taking video dataset. In this paper, we use an in-house dataset of five exam-taking and cheating-attempting videos.  Experimental results show that the proposed scheme has high precision and requires low computation cost. The proposed work can be considered as an initial work (which, to the best of our knowledge, is first such work) to address a new and practical research problem (i.e., privacy-preserving online proctoring). Further research is required to improve the results, e.g., using a large exam-taking video dataset.

Rest of this paper is organized as follows. Section~\ref{Sec:RelatedWork} discusses related work. In Section~\ref{Sec:PropWork}, we provide an overview of the proposed method. Section~\ref{Sec:FaceHiding} explains face hiding, and Section~\ref{Sec:AnDetect} explains anomaly detection. 
In Section~\ref{Sec:newResults}, we provide experimental results. Section~\ref{Sec:Conclusion} concludes and discusses future work.
\vspace{-0.2cm}

\section{Related Work}
\label{Sec:RelatedWork}

In the past, various cheating detection methods for online exams have been proposed. Some old ways of doing cheating detection are analysing students answer sheets for finding similarities  \cite{cleophas2021s} \cite{lemantara2018prototype}\cite{MANOHARAN2019139} or considering the exam takers previous academic performance \cite{souza2017conceptual}. Although these methods are sometimes useful \cite{souza2017conceptual}, they can be easily challenged as they do not video record the students. 

Recently, machine-learning-based  
online proctoring has been explored\cite{bateson2006cues,alessio2017examining,oskar2008online}. These methods can detect cheating by monitoring a student video from  the other side of the machine.  
Online proctoring methods generally fall under three categories: live proctoring, recorded proctoring, and automated proctoring \cite{hussein2020evaluation}. Live proctoring is a real-time proctoring system where a  human proctor monitors the exam-taking student through live videos, similar to the one shown in Figure \ref{fig:Example}. In live proctoring,  minimal information of the exam-taking student is recorded. On the other hand, recorded proctoring records the video and other log data of the exam taker. Later, such information is manually analysed for detecting cheating. Finally, automated proctoring aims to minimize the human involvement by replacing the human with an automated system. Automated proctoring is based on complex machine learning algorithms that can detect anomalies with the help of cloud-based high performance computers. Humans simply rely on the outcomes of the machine learning algorithms\cite{nigam2021systematic}. The automated proctoring is touted as the future for its cost-effectiveness and scalability. 

The list of private information that an exam taker gives up can be unanticipated and intrusive. Most of the above solutions have been built without considering student's privacy. Some of the examples of student's information that can be gathered during online exams are: audio, video (including 360\textdegree  panning video) , shared screen, keyboard strokes etc. \cite{hussein2020evaluation}.


To the best of our knowledge,  students privacy has not been addressed yet in an online proctoring system. In this paper, we show that proctoring is possible in a privacy-preserving manner. 
\vspace{-0.2cm}

\section{Proposed Method}
\label{Sec:PropWork}

   \begin{figure}[ht]
    \centering
    \includegraphics[width=0.45\textwidth]{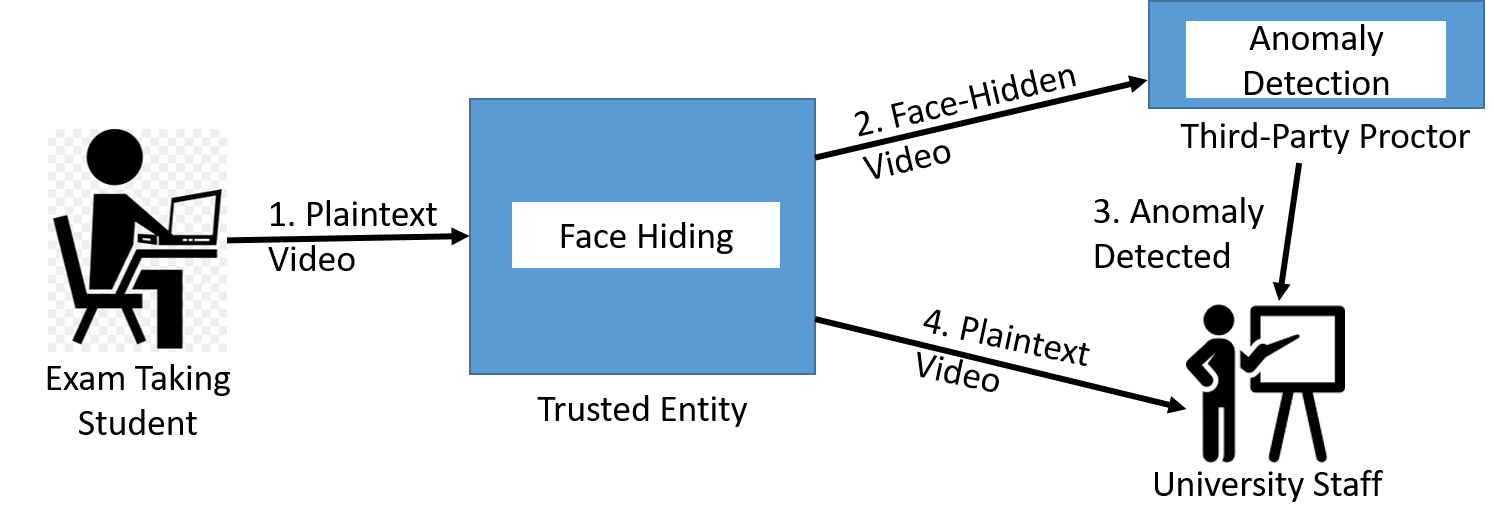}
    \caption{Architecture of the proposed system. Sequence of numbers 1,2,3,4 refers to the steps in the architecture of proposed system }
    \label{fig:Archi}
    \end{figure} 
Figure~\ref{fig:Archi} shows the overall architecture of the proposed system. There are four main players: a student, a trusted entity, an honest-but-curious third-party proctor, and a trusted university staff. We assume that the trusted entity can access the student's information, such as exam videos, photos, ID cards, in plaintext. This entity can either be present at the student-end (such as a trustzone in student's computing device) or at the university-end (such as a highly secure dedicated machine).  The communication between the student and the trusted entity is assumed to be secured. The third-party proctor is tasked to do the heavy-weight proctoring task. It is assumed that this entity will perform its task honestly but can be curious to know student's information, hence leading to privacy concerns. This entity can either be a third-party company hired by the university or an adhoc university department who has mostly crowd-sourced the proctoring activities. The trusted university staff can be a permanent staff who can access student's information in plaintext. This staff can be part of a small team (e.g., the team dealing with cheating) who will access the information of only those students flagged by the third-party proctor. 

The system will work as follows. A student will be asked to switch-on her webcam or selfie camera when she is taking the exam. Her live exam-taking video will be sent to the trusted entity in plaintext (Step 1). At the beginning, the trusted entity will verify student's ID. We assume that ID verification can be done by implementing an appropriate technique, such as the OCR technique \cite{baek2019wrong}~\cite{GitHubJa62:online}. The main job of the trusted entity is to hide facial information of the student. This is for minimizing privacy leaks. The face-hidden video will then be sent to the third-party proctor (Step 2). The third-party proctor will run anomaly detection on the face-hidden video for detecting potential cheating. This anomaly detection tool will serve as a triaging tool because of its light weight nature. Students flagged by this tool will then go through another round of manual check  at the third-party proctor-end (by viewing clipped anonymised video as shown in Figure~\ref{fig:anomaly_detected}). Those confirmed by the third-party proctor will be reported to the trusted university staff (Step 3). The university staff will finally get the clipped plaintext video from the trusted entity (Step 4) and take any further actions. 

In the following sections, the details of face hiding module and anomaly detection module will be discussed. These modules were developed using following datasets.

\subsection{Datasets}

\subsubsection{In-house dataset}

    
    \begin{figure}[hbt!]
    \centering
    \includegraphics[width=0.45\textwidth]{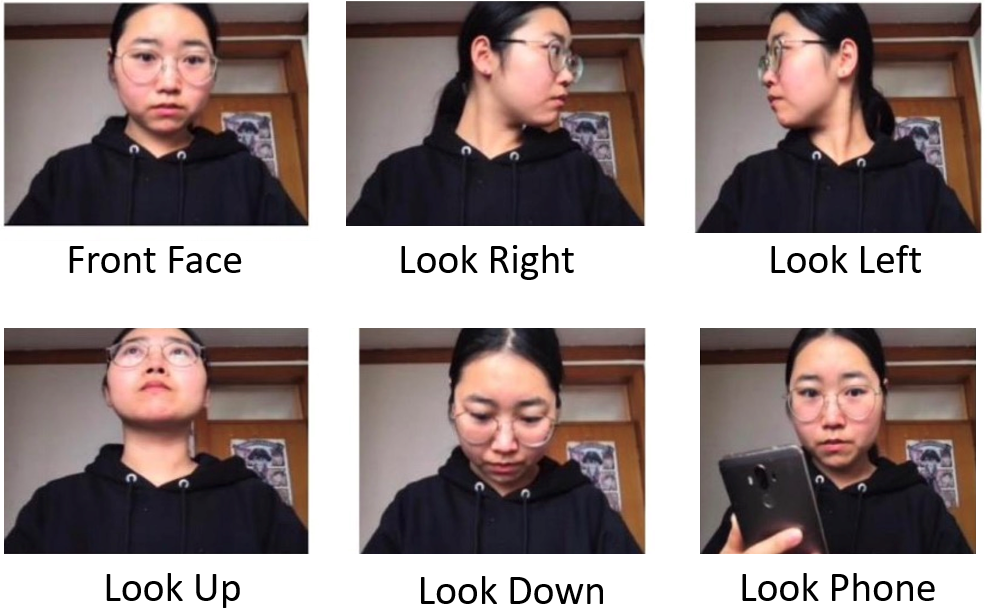}
    \caption{Various anomalies.}
    \label{fig:ano}
    \end{figure} 
    
The lack of a public exam-taking and cheating-attempting video dataset was one of the challenges of this project. Thus, we created an in-house dataset of three videos taken from five different Asian-origin $25$ to $30$ years old participants. Each video is roughly two to three minute of a mock-up exam, with $1280\times720$ resolutions  and $25$ to $30$ FPS (frames per second).  The cheating was attempted by creating an anomaly as shown in Figure~\ref{fig:ano}. We  assume that in a normal exam-taking condition, a student will look at  the front computer screen or the keyboard. Any other movements or gestures from the upper body can be an anomaly. Figure~\ref{fig:ano} shows various poses a student can create when taking the exam. The \textit{Front Face} pose represents the normal exam-taking behaviour. Other poses show typical student behaviour when she is trying to cheat, e.g., when peeking on a cheat sheet or flipping a book. 

\subsubsection{Public Datasets}
\label{modulemethod}
Because the in-house dataset is small, we reconfirmed the results in Section~\ref{Sec:FaceHiding}  using the following public image datasets:  HELEN \cite{HELEN}, UTKFace \cite{UTKFace}, CelebA  \cite{CelebA}, RF \cite{RF}, and LFW \cite{LFW}. The public dataset helped to asses the privacy leak from anonymised faces.

\section{Face Hiding}
\label{Sec:FaceHiding}

    \begin{figure}[ht]
    \centering
    \includegraphics[width=0.5\textwidth]{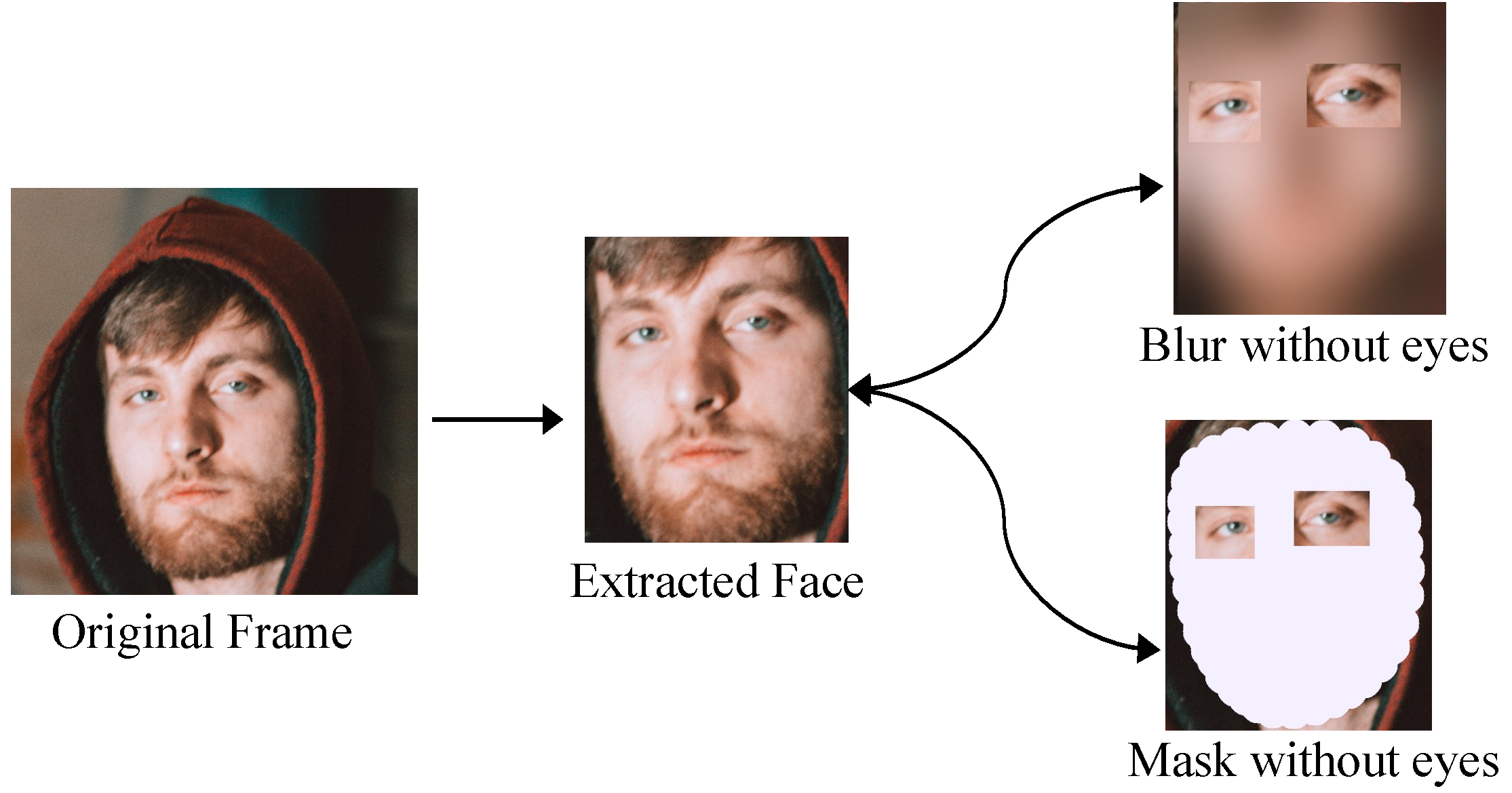}
    \caption{Frame-by-frame blurring or masking applied on public dataset.}
    \label{fig:6_figture_11}
    \end{figure} 

\begin{figure}[hbt!]
    \centering
    \includegraphics[width=0.4\textwidth]{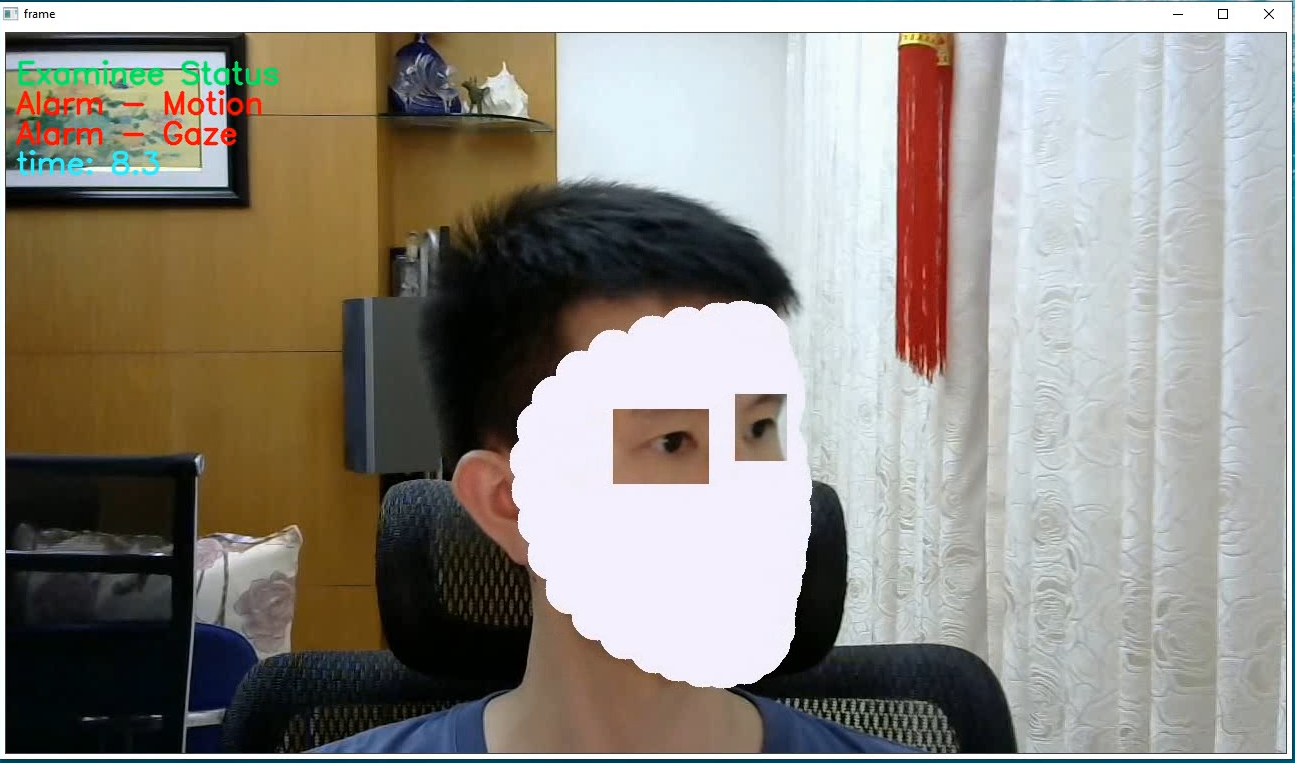}
    \caption{Anomaly detected under masked video.}
    \label{fig:anomaly_detected}
    \end{figure}

The Face Hiding module hides the student's facial information from a video for minimizing privacy leak. We have used blurring or masking to hide the face (Figure~\ref{fig:6_figture_11}). In the hidden face, the eyes are not hidden for facilitating anomaly detection (i.e., cheating detection) as shown in Figure~\ref{fig:anomaly_detected}. At first, the face and eyes are detected, and then blurring or masking is applied. 
\subsection{Face and Eye Detection}
Machine learning-based approach has been used to find face and eyes from a given exam taking video sequentialy (frame-by-frame). A number of pre-trained machine-learning models are available for detecting a face and an eye from a frame (aka image). 
We compared Dlib and MediaPipe models\footnote{We compared range of machine learning models and selected the Dlib and MediaPipe. Due to page limitation, we are reporting the best two.} as they do not require high-performance computers and give highest detection rate (based on our experiment with in-house dataset as discussed below). They were run on a local Windows $10$ computer having $16$ GB RAM and i7-10710U CPU. Face detection and eyes detection results were compared individually. The in-house dataset was first used.  Then HELEN dataset was used as it has images having similar resolution to video frames. 

\subsubsection{Results using in-house dataset}

\textit{Face detection result:} All the models performance were compared using $detected\ rate = \frac{Number\ of\ faces\ detected}{Actual\ number\ of\ faces} \times 100 $ metric.  For this, the frames of test videos were first manually labeled as \textit{face} and \textit{no face} for getting the ground truth (which is called as \textit{Actual number of faces} in the formula). \textit{Eye detection result:} The eye detection rates were also obtained by first labeling the videos and then running the models. The detection rate was computed as $detected\ rate = \frac{Number\ of\ frames\ with\ correct\ eyes\ detection}{Number\ of\ frames\ labeled\ as\ ``face"}$. The detection results of the models are given in Table~\ref{tab:6_table_41}.




We then performed our experiment using the HELEN dataset. Note that for the MediaPipe model, the eye detection rate is different than the face detection rates (unlike Dlib). This is because the MediaPipe model detects face  and  eyes independently. 
    

\subsubsection{Results using HELEN dataset}
Table~\ref{tab:6_table_41} show the detection results of both the models individually and also when they are combined (hybrid mode). Based on the results, the hybrid mode was chosen.

\begin{figure}[ht]
    \centering
    \includegraphics[width=0.45\textwidth]{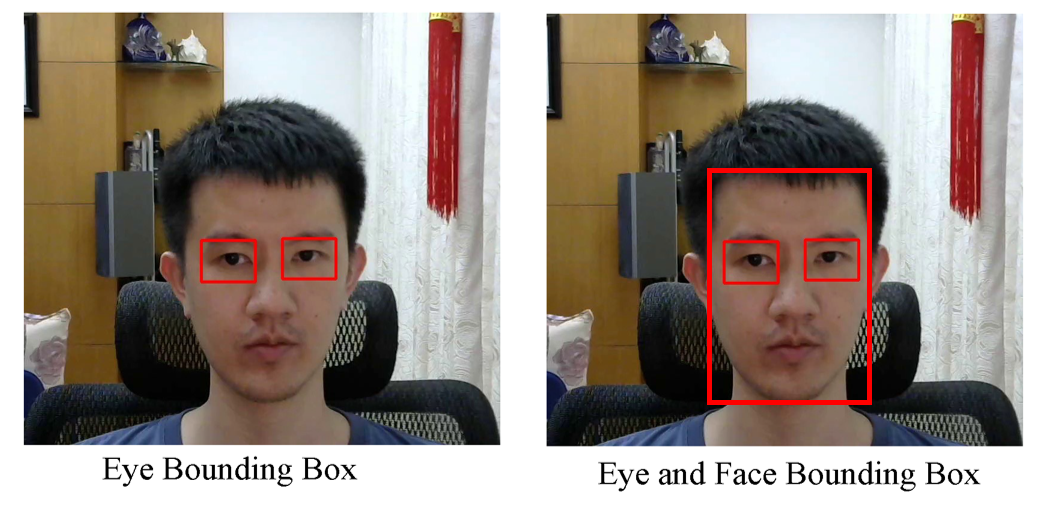}
    \caption{Bounding boxes.}
    \label{fig:5_figure_21}
    \end{figure}

First, the MediaPipe is used because of its higher FPS rate. If MediaPipe fails, then Dlib is used.
Based on our experiment, we found that this arrangement can detect face and eyes with a very high precision rate. Thus, if no face and eyes were found, we assumed that the input frame did not contain face and eyes. In that case, the face and eyes of the previous frame were considered. As show in  Figure~\ref{fig:5_figure_21}, bounding boxes representing the detected face and eyes were output to the next step.  

\begin{table}[ht]
 \caption{Face and eye detection rate using in-house and HELEN dataset.}
    \centering
    \begin{tabular}{llll}
      & \textbf{Dlib} & \textbf{MediaPipe} & \textbf{Hybrid}\\ \midrule
    \textbf{Face Detected Rate (in-house data)} & 92.93\% & 91.44\% & 95.3\%\\  
    \midrule
    \textbf{Face Detected Rate (HELEN data)} & 96.93\% & 78.67\% & 99.27\%\\ 
    \textbf{Average Normalized Error} & 2.65 & 4.30 & 3.99\\ 
    \midrule
    \textbf{Eyes Detected Rate (in-house data)} & 92.93\% & 97.41\% & 97.8\%\\  
    \midrule
    \textbf{Eyes Detected Rate (HELEN data)} & 96.93\% & 78.67\% & 99.27\%\\ 
    \textbf{Average Normalized Error} & 0.72 & 4.27 & 3.55\\ 
    \end{tabular}
    \label{tab:6_table_41}
    \end{table}

    
    \begin{figure}[hbt!]
    \centering
     \includegraphics[width=0.45\textwidth]{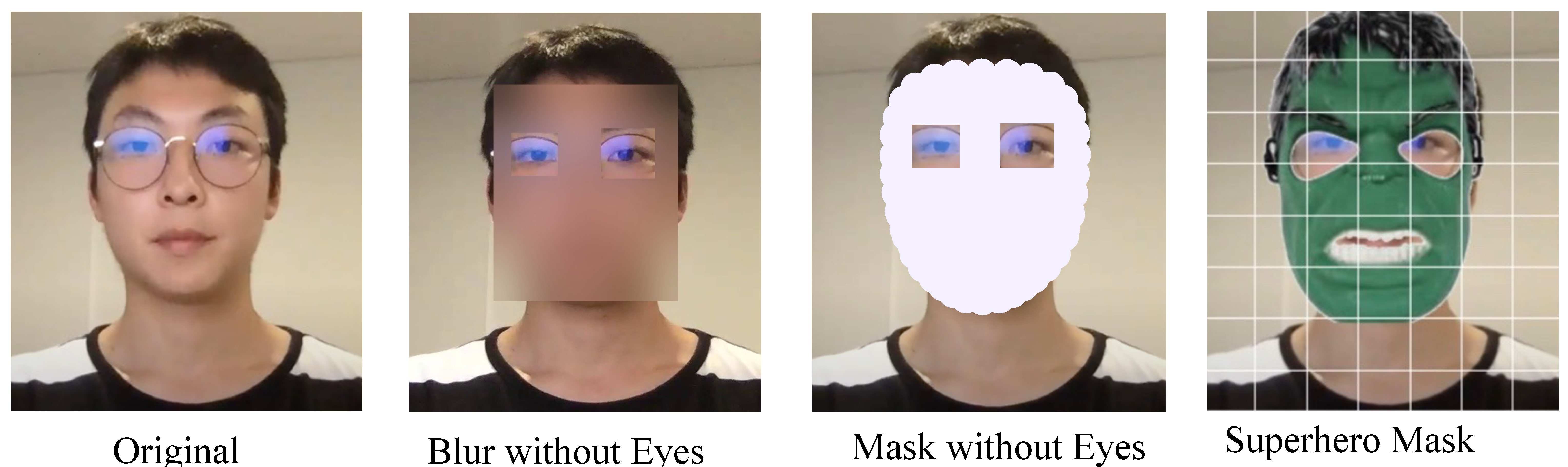}
    \caption{Different face hiding methods.}
    \label{fig:5_table_1}
    \end{figure}

\begin{table}[ht]
    \centering
    \caption{Recognition rate and speed performance on public datasets.}
    \begin{tabular}{lllll}
     \textbf{Method} & \textbf{UTKFace} & \textbf{RF} & \textbf{Celebahq} & \textbf{LFW}  \\ 
     \midrule
    \textbf{Blur w/o Eyes} & 28.43\% & 0.63\% & 2.29\% & 0.71\% \\
    \textbf{Mask w/o Eyes} & 0.14\% & 0\% & 0\% & 0\%\\ 
    \end{tabular}
    \label{tab:6_table_7}
    \end{table}
    
    
\subsection{Assessment of face hiding module} After the final schema of the privacy-preserving part and the blur level are decided, the effect of privacy-preserving methods will be evaluated using public datasets. Either blurring or masking is used to hide the facial information except the eyes. Blurring is a widely used method \cite{zhangfacial}. Therefore, although we found that the masking is better than blurring in preserving privacy (as shown in  Table~\ref{tab:6_table_7}) and the computation cost blurring was over hundred times more than that of masking, $63$ and $28,852$ FPS respectively. For sake of completeness we have also conducted experiments with blurring.  Guassian blurring with blur level $30$ has been used in this paper. A single white mask (\textit{Mask without Eyes} in Figure~\ref{fig:5_table_1}) has  been considered for masking the frames of all videos.   For this, first 3D face landmarks (key facial features) are detected (MediaPipe returns $468$  3D  face landmarks and  Dlib returns $68$ face landmarks), and then these landmarks are blocked using point size $26$. Note that different masks can also be used for different students and this mask can also be  personalised based on user preferences (e.g., using Superman mask for one student and Hulk mask for another student, as shown in Figure~\ref{fig:5_table_1}). 

\section{Anomaly Detection}
\label{Sec:AnDetect}
The objective of the anomaly detection module is to detect anomalies even when the student's face is blurred or masked. 
Image hashing-based anomaly detection method has been used.

\subsection{Image Hashing Method}
Traditionally, image hashing\cite{phash}, or perceptual hashing, is used for finding similar images. When the difference in hash values of two images is less than a threshold, the images are considered as similar. In this paper, we have used this phenomena for finding if all the frames of an exam-taking video are similar or there are some frames different than others.  When the frames are similar, they supposedly have the same pose position. If different, they show at least two different pose positions, and hence the possibility of an anomaly. The hash comparison is done by comparing each subsequent frame with an anchor frame. Hamming distance\cite{norouzi2012hamming} is used for finding the difference in the hash values. A previous study has found that the dHash hashing method is faster and has better detection rate than other two popular hashing methods, aHash and pHash \cite{dhash}. Our experiment with public datasets also confirmed this claim. We therefore have used the dHash hashing method with hash size $12$. 




 \subsubsection{Determining threshold}


Setting the threshold is tricky as the hash difference depends on the exam-taking environment. For example, if a student is closer to the camera, the hash difference will be larger than the hash difference obtained when the student is far from the camera. Thus, we use a tailored threshold for each exam-taking session, which is computed from the student's sitting pattern at the beginning of the exam.

\begin{figure}[ht]
    \centering
    \makebox[0.45\textwidth][c]{
    \includegraphics[width=0.45\textwidth]{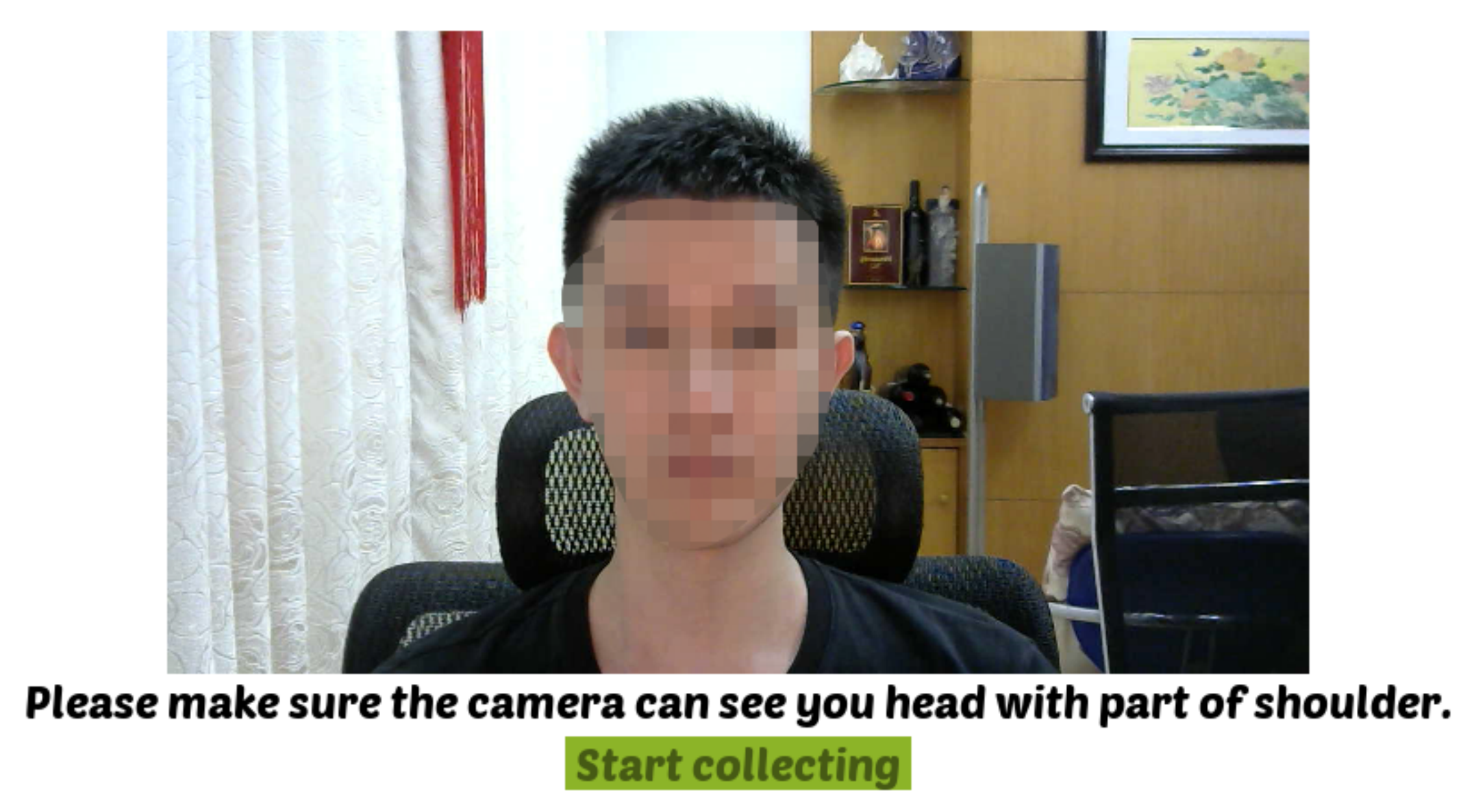}
    }
    \caption{Instruction to adjust sitting position.}
    \label{fig:gaze_instruction}
\end{figure}

\begin{figure}[ht]
    \centering
    \makebox[0.45\textwidth][c]{
    \includegraphics[width=0.45\textwidth]{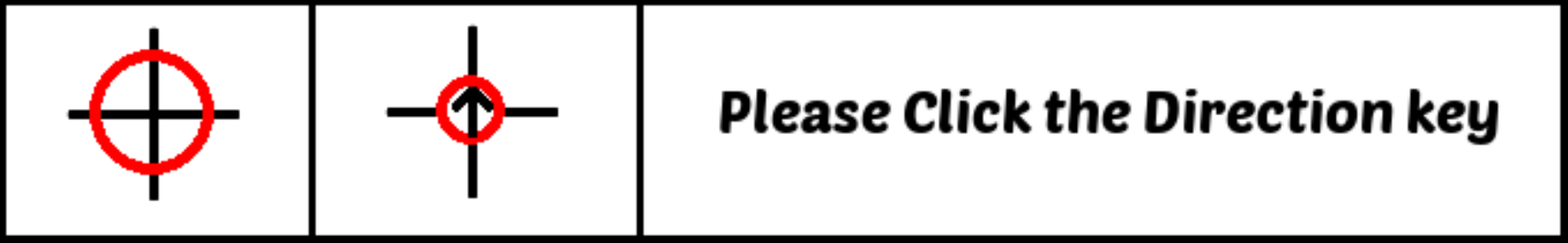}
    }
    \caption{Shown symbols}
    \label{fig:gaze_instruction2}
\end{figure} 

At the beginning, student's sitting pattern is recorded by her sitting position and eye interaction with the computer. The correct sitting position is set by showing a real-time camera view to the student, and asking the student to adjust the sitting position accordingly (Figure~\ref{fig:gaze_instruction}). A method similar to Krafka et al.'s proposed method \cite{itracker}' is then used to record the eye interaction reliably. The core idea of Krafka et al's method is to show a random symbol at several screen positions, and ensure that the student gazed them. For this, we show the student a crosshair and circle pair at random screen positions (Figure~\ref{fig:gaze_instruction2}). The circle shrinks to the center of the crosshair to guide the student where to look. Once the circle shrunk to the minimum size, a random arrow out of ($\uparrow, \downarrow, \rightarrow, \leftarrow$) will show up for $0.5$ seconds. During this $0.5$ seconds, the computer camera takes a photo of the student. Then the student is asked to click the corresponding direction button on the keyboard to save the photo (after $0.5$ seconds). If the student responds correctly, the photo and corresponding symbol coordinates  are saved, and the crosshair and circle pair is shown in a new screen position. Otherwise, the pair is shown in the same screen position. 


We assume that the above sitting pattern covers all interactions a student will do with the computer when siting with her normal exam-taking pose. The collected photos, therefore, are used to obtain the threshold. The difference in the hash value of one photo with the hash value of all other photos are obtained, and the maximum hash difference is set as the threshold. 

\subsubsection{Selecting the anchor frame}
    \begin{figure}[hbt!]
    \centering
    \includegraphics[width=0.5\textwidth]{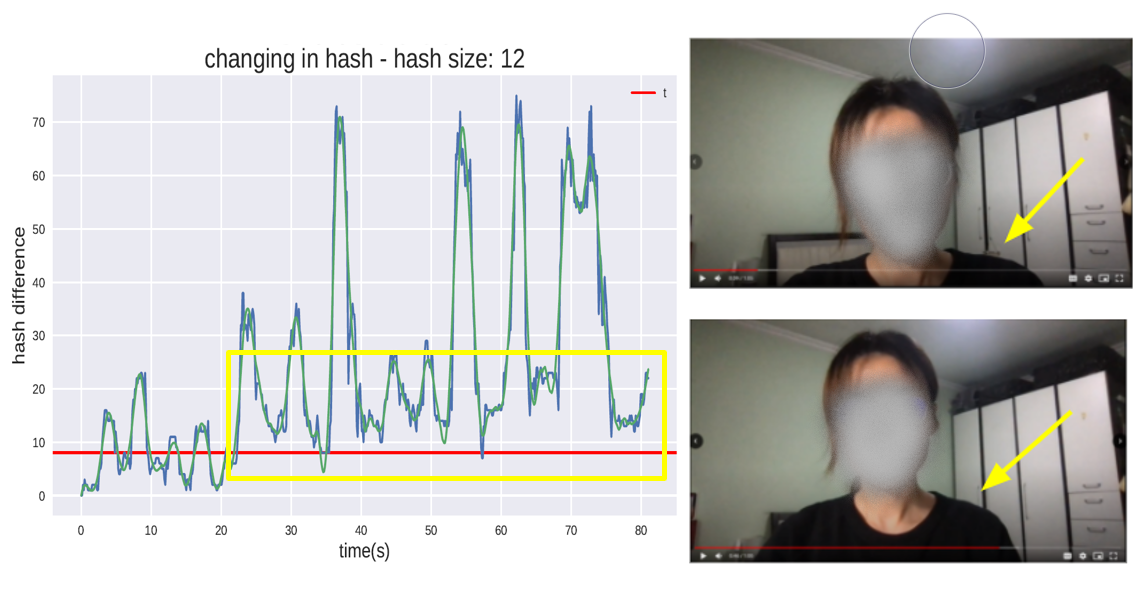}
    \caption{Illustrates the issue when a fixed anchor frame is used to compute hash difference.}
    \label{fig:hash-ref2}
    \end{figure} 
A simple approach is to set the first frame of the exam-taking video  as the anchor frame, and compare all other frames with it. As illustrated in Figure~\ref{fig:hash-ref2}, this approach, however, does not work well when the student changes her sitting position during the exam. In this example, the student has two sitting positions: closer to camera position, and far from the camera position (both shown using yellow arrows). Initially, the student was closer to the camera. The hash difference in this case is a bit lower. But, when the student sits a bit far from the camera, the hash difference becomes significantly higher (highlighted using yellow box in the graph). Such significant change in hash values can lead to poor performance. This issue can be addressed by  either adjusting the threshold 
or changing the anchor frame when the student changes her position. Adjusting the threshold will need that the student goes through the above calibration process once again in the middle of the exam (which is clearly very user unfriendly). Changing the anchor frame does not have such requirement. Thus, we used this method. In this case, the threshold found from the initial calibration is used through out the exam session. 

    
    \begin{figure}[hbt!]
    \centering
    \includegraphics[width=0.5\textwidth]{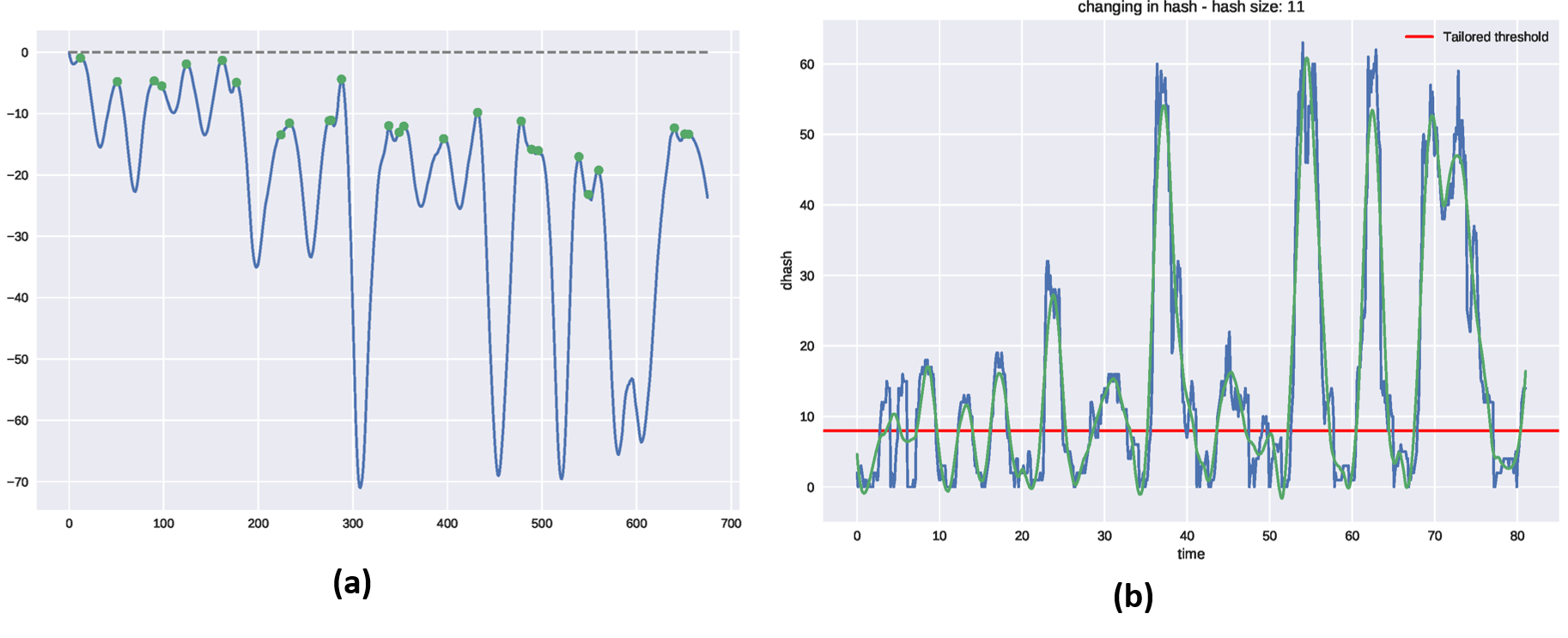}
    \caption{(a): How a new anchor frame is found is illustrated. (b): Image hashing is re-scanned using the new anchor frame.}
     \label{fig:hash-peak}
    \end{figure}

Ideally, the anchor frame 
can be changed when the student changes her sitting position. A changed position can be identified by analysing the student's face in consecutive frames of the video. When a student sits closer to the screen, her face will be bigger on the frame than when she is far from the screen. In this paper, we, however, use a different approach for changing the anchor frame. We change the anchor frame whenever a potential normal behaviour is found after potential anomalous behaviour. The normal behaviour can be found by analysing the plot in Figure~\ref{fig:hash-ref2}. In this plot, the valley points represent the potential normal behaviours. This is because the student will look at the computer screen (for typing the result) even after looking away from it for sometime (e.g., when reading a book), therefore creating the valley points. As part of the analysis, we first apply a widely used  Savitzky–Golay filter\cite{guest2012numerical} for smoothing (removing any noise) the signal presented in the plot. The smoothed plot is then inverted (as shown in Figure~\ref{fig:hash-peak} (a)). The frames corresponding to the peak points of the inverted-plot (shown as green dots in the plot) are considered as the anchor frames. Each time a new green dot is found, the anchor frame is reset to the frame corresponding to this dot, and all future hash difference calculations are done using the new anchor frame. Figure~\ref{fig:hash-peak} (b) shows the plot using such anchor frames.

\vspace{-0.2cm}

\section{Results}
\label{Sec:newResults}
The experiment was done using an in-house test dataset of three exam-taking videos collected from three different participants. Each video was two to three minutes long. Detailed instruction of how to emulate an exam and attempt cheating was provided to each participant. Each frame of a video was then manually labeled as either a normal pose or an anomaly pose. 

MediaPipe and Dlib-based face and eye detection, Gaussian blurring and  single-white masking, and dHashing-based image hashing with hash size $12$ were used. The experiment was done on a  local  Windows 10 computer  having 16 GB  RAM and  i7-10710U  CPU. Each video frame was first processed by MediaPipe and Dlib (for detecting face and eye) module, then by blurring or masking-based face hiding module, and finally by the dHashing-based image hashing module. The obtained anomaly results were compared with the ground truth. 

\begin{table}[h]
    \centering
    \caption{Performance with two different privacy preserving modes (Acc. for Accuracy, Rec. for Recall, Pre. for Precision).}
    \begin{tabular}{llllllll}
     & Blur & & & Mask & \\ \midrule
     Participant & Acc. & Rec. & Pre. & Acc. & Rec. & Pre.\\ \midrule
   1 & 67.4\% & 41.2\% & 83.3\% & 76.6\% & 64.7\% & 83.3\% \\
   2 & 76.6\% & 54.1\% & 95.8\% & 77.7\% & 61.2\% & 89.7\% \\
   3 & 78.2\% & 60.0\% & 92.3\% & 80.6\% & 63.8\% & 94.4\% \\
    \end{tabular}
    \label{tab:7_system_test_tab}
\end{table}



%

Table~\ref{tab:7_system_test_tab} shows the experimental result. The result is different for different participants as they acted differently for creating the same anomaly behaviour. For example, when asked to look left, one looked to a bit more left than the other. 
The low recall rate is due to the fact that only looking at the screen was considered as normal behaviour, and even looking at the edge of the screen was labeled as anomaly.  

The proposed model has low computation overhead. The average running of blurring-based and masking-based methods  measured in terms of FPS are 31 and 35 FPS, respectively. As expected, masking approach performed better, although both the approaches can be easily performed on normal PCs. Further speed performance could be improved by employing cropping and simple scaling-based approach \cite{yaqub2018towards}. Moreover, the proposed module could easily be adopted to existing proctoring systems as well as zoom-based proctoring \cite{kujathnovel} with added level of security \cite{mohanty2020seamless}

\vspace{-0.2cm}

\section{Conclusion and Future Work}
\label{Sec:Conclusion}
Online student proctoring is a reality in online-based exams. In this paper, we proposed a privacy-preserving online proctoring system using  image-hashing-based anomaly detection. Experiments showed promising results.

There are a several ways how this preliminary work can be further improved. The first and the far most requirement is creating a large exam-taking students dataset.  Secondly, the proposed method can be improved by exploring other privacy-preserving measures and by considering on other anomalies.  
\vspace{-0.4cm}
\section*{Acknowledgement}
We would like to thank Jason, Yang, Mill, Doris, Vivi for the implementation and testing of the algorithms on their dataset.

\bibliographystyle{IEEEtran}

\bibliography{Bib1}

\end{document}